# Thermodynamic stability of new phase $Bi_{12.5}Sm_{1.5}ReO_{24.5}$

N.I. Matskevich, Th. Wolf, A.N. Bryzgalova, T.I. Chupakhina, E.S. Zolotova, M.Yu. Matskevich, M.A. Bespytov

ABSTRACT

The $Bi_{12.5}Sm_{1.5}ReO_{24.5}$ phase has been synthesized for the first time. The preparation of $Bi_{12.5}Sm_{1.5}ReO_{24.5}$ has been performed by solid-state reaction from $Bi_2O_3$, $Gd_2O_3$, $Re_2O_7$. The X-ray measurements have showed that $Bi_{12.5}Sm_{1.5}ReO_{24.5}$ was cubic structure (space group *Fm3m*). The standard molar enthalpy of formation of $Bi_{12.5}Sm_{1.5}ReO_{24.5}$ has been determined by solution calorimetry combining the enthalpies of dissolution of $Bi_{12.5}Sm_{1.5}ReO_{24.5}$ and $6.25Bi_2O_3$ + $0.75Sm_2O_3$ + $0.5Re_2O_7$ mixture in 2 M HCl and literature data. It has been obtained that above-mentioned phase is thermodynamically stable with respect to their decomposition into binary oxides at room temperatures.

## 1. Introduction

Compounds on the basis of $\delta$-$Bi_2O_3$ are perspective ionic conductors [1-10]. The oxygen deficient fluorite structure of $\delta$-$Bi_2O_3$ provides exceptionally high oxide ion conductivity: 1 S $cm^{-1}$ at 750 °C, 2-3 orders of magnitude greater than that of Ca- or Y-stabilized zirconias. Six polymorphs of this oxide, labeled the $\alpha$-, $\beta$-, $\gamma$-, $\delta$-, $\varepsilon$-, and $\omega$-phases, have been reported in the literature. The room temperature stable variety of $Bi_2O_3$ is the $\alpha$ monoclinic form. On heating, the $\alpha$-variety transforms into the $\delta$-form; various transition temperatures ranging from 717 to 740 °C were indicated until the value 730(1) °C was generally accepted. The high temperature form $\alpha$-$Bi_2O_3$ is stable between 730 °C and approximately 825 °C, the melting temperature of the material.

There were a lot of attempts to stabilize $\delta$- form of $Bi_2O_3$ in literature. Stabilizing $\delta$ -$Bi_2O_3$ with the smaller rare earth cations has been studied widely, and 20% Er substitution provides conductivities at 530 °C that are comparable to stabilized zirconias at 830 °C. Wide domains of solid solutions of this variety have been obtained by doping with one or several aliovalent cations (2+ to 6+), and they constitute a series of materials with a high ionic conduction level at moderate temperatures. Authors of paper [1] reported the conductivity and structural data on some new materials ($Bi_{12.5}La_{1.5}ReO_{24.5}$, $Bi_{12.5}Nd_{1.5}ReO_{24.5}$, $Bi_{12.5}Eu_{1.5}ReO_{24.5}$ , $Bi_{12.5}Y_{1.5}ReO_{24.5}$ that are related to cubic δ-$Bi_2O_3$ but have low-temperature (<400 °C) oxide ion conductivities that are significantly higher than previously reported for δ-$Bi_2O_3$ phases and comparable to those of BIMEVOX materials.

Here we report synthesis, structure and thermochemical investigation of new phase with the composition $Bi_{12.5}Sm_{1.5}ReO_{24.5}$.

## 2. Experimental

*Synthesis of samples*

Synthesis of $Bi_{12.5}Sm_{1.5}ReO_{24.5}$ was performed by solid state reaction:

$6.25\ Bi_2O_3 + 0.5Re_2O_7 + 0.75\ Sm_2O_3 = Bi_{12.5}Sm_{1.5}ReO_{24.5}$.

Polycrystalline samples of $Bi_{12.5}Sm_{1.5}ReO_{24.5}$ were prepared from bismuth oxide, samarium oxide and rhenium oxide. Before synthesis $Sm_2O_3$ was annealed at 1070 K up to constant weight. We used the following reagents $Sm_2O_3$ (99.99%, Johnson Matthey Company), $Bi_2O_3$(99.999%, ABCR, Karlsruhe), $Re_2O_7$(99.99%, Alfa).

Stoichiometric amounts of $Bi_2O_3$, $Sm_2O_3$, $Re_2O_7$ were mixed in an agate container with agate balls using a planetary mill (FRITSCH pulverisette) for 72 h with 8 intermediate regrounds. All operations with $Re_2O_7$ were performed in a glove box under pure argon, because $Re_2O_7$ reacts with water. The ground materials were pelletized using a 10-mm-diameter die and fired at 1070 K in air for 70 h.

Phase purity and composition of $Bi_{12.5}ReSm_{1.5}O_{24.5}$ was confirmed by X-ray power diffraction (XPD) using a STADI-P, Stoe diffractometer, using Cu-$K_{\alpha 1}$ radiation and a WDX-spectrometer (ARL ADVANT'XP). The samples were shown to be phase-pure ceramics with a cubic structure (space group *Fm3m*). The refined cell parameter and cell volume obtained for $Bi_{12.5}$ $Sm_{1.5}ReO_{24.5}$ are 5.5958 (1) *Å*, 175.227(6) $Å^3$ respectively. Detail structural information is presented in Table 1. X-ray data were treated in FULLPROF-2010. δ-$Bi_2O_3$ was used as base line. The power X-ray diffraction pattern of bismuth perrhenate doped by samarium oxide is presented in Figure 1. We compared X-ray data for $Bi_{12.5}Sm_{1.5}ReO_{24.5}$ with X-ray data for $Bi_2O_3$ [11], $Re_2O_7$ [12], $Sm_2O_3$ [13] and found that there were no reflexes of $Bi_2O_3$, $Re_2O_7$, $Sm_2O_3$ on our X-ray diagram.

*Calorimetric Technique*

The calorimetric experiments were performed in an automatic dissolution calorimeter with an isothermal shield. The main part of the calorimeter was a Dewar vessel with a brass cover (V = 200 ml). The platinum thermometer, calibration heater, cooler, mixer, and device to break the ampoules were mounted on the lid closing the Dewar vessel. The construction of the solution calorimeter and the experimental procedure are described elsewhere [4]. Dissolution of potassium chloride in water was performed to calibrate the calorimeter. The obtained dissolution heat of KCl was 17.41 ± 0.04 kJ $mol^{-1}$ (the molality of the final solution was 0.028 mol $kg^{-1}$, T = 298.15 K). The literature data are: 17.42 ± 0.02 kJ $mol^{-1}$), 17.47 ± 0.07 kJ $mol^{-1}$ [14, 15].
The experiments were performed at 298.15 K. The amount of substance used was about 0.2 g.

*Thermochemical Cycles*

The thermochemical cycle for determination of formation enthalpy of $Bi_{12.5}Sm_{1.5}ReO_{24.5}$ was constructed in such a way that bithmuth oxide and phases of $Bi_{12.5}Sm_{1.5}ReO_{24.5}$ were dissolved in 2 M HCl. The samarium oxide was dissolved in solution 2 (2 M HCl with dissolved $Bi_2O_3$) and rhenium oxide was dissolved in solution 3 (2 M HCl with dissolved $Bi_2O_3$ and samarium oxide). Then the enthalpy of dissolution of mixture $Bi_2O_3$, $Sm_2O_3$, $Re_2O_7$ was compared with the solution enthalpy of $Bi_{12.5}Sm_{1.5}ReO_{24.5}$. The reactions showing the principal scheme from which the formation enthalpies from binary oxides are calculated are presented below:

$$6.25\ Bi_2O_3(s) + \text{solution 1} = \text{solution 2} + 6.25\ \Delta_{sol}H^o\ (Bi_2O_3(s)) \quad (1)$$

$$0.75\ Sm_2O_3(s) + \text{solution 2} = \text{solution 3} + 0.75\Delta_{sol}H^o\ (R_2O_3(s)) \quad (2)$$

$$0.5\ Re_2O_7(s) + \text{solution 3} = \text{solution 4} + 0.5\Delta_{sol}H^o\ (Re_2O_7(s)) \quad (3)$$

$Bi_{12.5}Sm_{1.5}ReO_{24.5}(s)$ + solution 1 = solution 4 + $\Delta_{sol}H^o(Bi_{12.5}Sm_{1.5}ReO_{24.5}(s))$ (4)

Here, solution 1 is 2 M hydrochloric acid.

The measured enthalpies of dissolution (1)-(4) were used for calculating the enthalpy of the reaction:

$$6.25\ Bi_2O_3 + 0.75Sm_2O_3 + 0.5Re_2O_7 = Bi_{12.5}Sm_{1.5}ReO_{24.5} + \Delta_{ox}H^o \quad (5)$$

as following:

$$\Delta_{ox}H^o = 6.25\ \Delta_{sol}H^o\ (Bi_2O_3(s)) + 0.75\Delta_{sol}H^o\ (R_2O_3(s)) + 0.5\Delta_{sol}H^o\ (Re_2O_7(s)) -$$

$$\Delta_{sol}H^o(Bi_{12.5}Sm_{1.5}ReO_{24.5}(s)),$$

where $\Delta_{ox}H^o$ is the formation enthalpy of $Bi_{12.5}Sm_{1.5}ReO_{24.5}$ from binary oxides.

**3. Results and discussion**

To determine the formation enthalpy of $Bi_{12.5}Sm_{1.5}ReO_{24.5}$ we measure the solution enthalpies of bismuth oxide, samarium oxide, and rhenium oxide and solution enthalpy of $Bi_{12.5}Sm_{1.5}ReO_{24.5}$. The dissolution enthalpies for binary oxides are presented in Table 1 together with data on enthalpies of formation, which were taken from [4]. The solution enthalpy of investigated compound is $\Delta_{sol}H°(Bi_{12.5}Sm_{1.5}ReO_{24.5}\ (s)) = -882.8 \pm 3.7$ kJ/mol. All experimental values were calculated as the average value from six parallel experiments. The errors were calculated for 95% confidential interval using standard procedure of treatment of experimental results [14].

The measured enthalpies of dissolution were used for calculating the enthalpy of $Bi_{12.5}Sm_{1.5}ReO_{24.5}$ from binary oxides:

$$6.25\ Bi_2O_3 + 0.75\ Sm_2O_3 + 0.5\ Re_2O_7 = Bi_{12.5}Sm_{1.5}ReO_{24.5} + \Delta_{ox}H°_5$$

as follows: $\Delta_{ox}H°_5 = 6.25\Delta_{sol}H°_1 + 0.75\Delta_{sol}H°_2 + 0.5\Delta_{sol}H°_3 - \Delta_{sol}H°_4$

Then, we calculated the standard formation enthalpy according to formula:

$$\Delta_f H°(Bi_{12.5}Sm_{1.5}ReO_{24.5}) = \Delta_{ox}H°_5 + 6.25\Delta_f H°(Bi_2O_3) + 0.75\Delta_f H°(Sm_2O_3) + 0.5\Delta_f H°(Re_2O_7).$$

In this way, we obtained the values for formation from binary oxides and standard formation enthalpy of $Bi_{12.5}Sm_{1.5}Re\,O_{24.5}$ as following:

$\Delta_{ox}H° = –141.3 \pm 5.6$ kJ/mol – the formation enthalpy from binary oxides;

$\Delta_f H° = –5760.1 \pm 8.2$ kJ/mol – the standard formation enthalpy.

As a conclusion, it was established that the compound is thermodynamically stable with respect to the decomposition to binary oxides. It is also interesting to study the thermodynamic stability of the $Bi_{12.5}Sm_{1.5}ReO_{24.5}$ compounds with respect to mixtures including other stable ternary oxides in subsystems $Bi_2O_3$-$Sm_2O_3$, $Bi_2O_3$-$Re_2O_7$, $Sm_2O_3$-$Re_2O_7$ but there are no thermodynamic data available now on these phases. It is our task for future.

**Conclusions**

1. For the first time the enthalpies of solution of $Bi_{12.5}Sm_{1.5}ReO_{24.5}$ were measured using 2 M HCl at 298.15 K.

2. Using measured solution enthalpies of $Bi_{12.5}Sm_{1.5}ReO_{24.5}$, $Bi_2O_3$, $Sm_2O_3$, $Re_2O_7$ and literature data on standard enthalpies of $Bi_2O_3$, $Sm_2O_3$, $Re_2O_7$ the standard formation enthalpy of $Bi_{12.5}Sm_{1.5}ReO_{24.5}$ and formation enthalpy from binary oxides were calculated for the first time. It was established that $Bi_{12.5}Sm_{1.5}ReO_{24.5}$ is thermodynamically stable with respect to decomposition to binary oxides.

**Acknowledgment**

This work is supported by Karlsruhe Institute of Technology (Germany), RFBR (project 13-08-00169) and Program of Fundamental Investigation of Siberian Branch of the Russian Academy of Sciences.

Table 1. Structural parameters of $Bi_{12.5}Sm_{1.5}ReO_{24.5}$

| | |
|---|---|
| Compound | $Bi_{12.5}Sm_{1.5}ReO_{24.5}$ |
| Crystal system | Cubic |
| Space group | Fm-3m |
| Lattice parameters<br>a<br>cell volume | <br>5.5958 (1) *Å*<br>175.227(6) $Å^3$ |
| O2<br>z | <br>0.369 (3) |
| $R_{wp}$<br>$R_B$<br>$R_f$ | 3.21<br>2.33<br>2.20 |

Atomic fractional coordinates of (Bi/Ln/Re) are (0,0,0), those O(1) are (1/4, 1/4, 1/4)

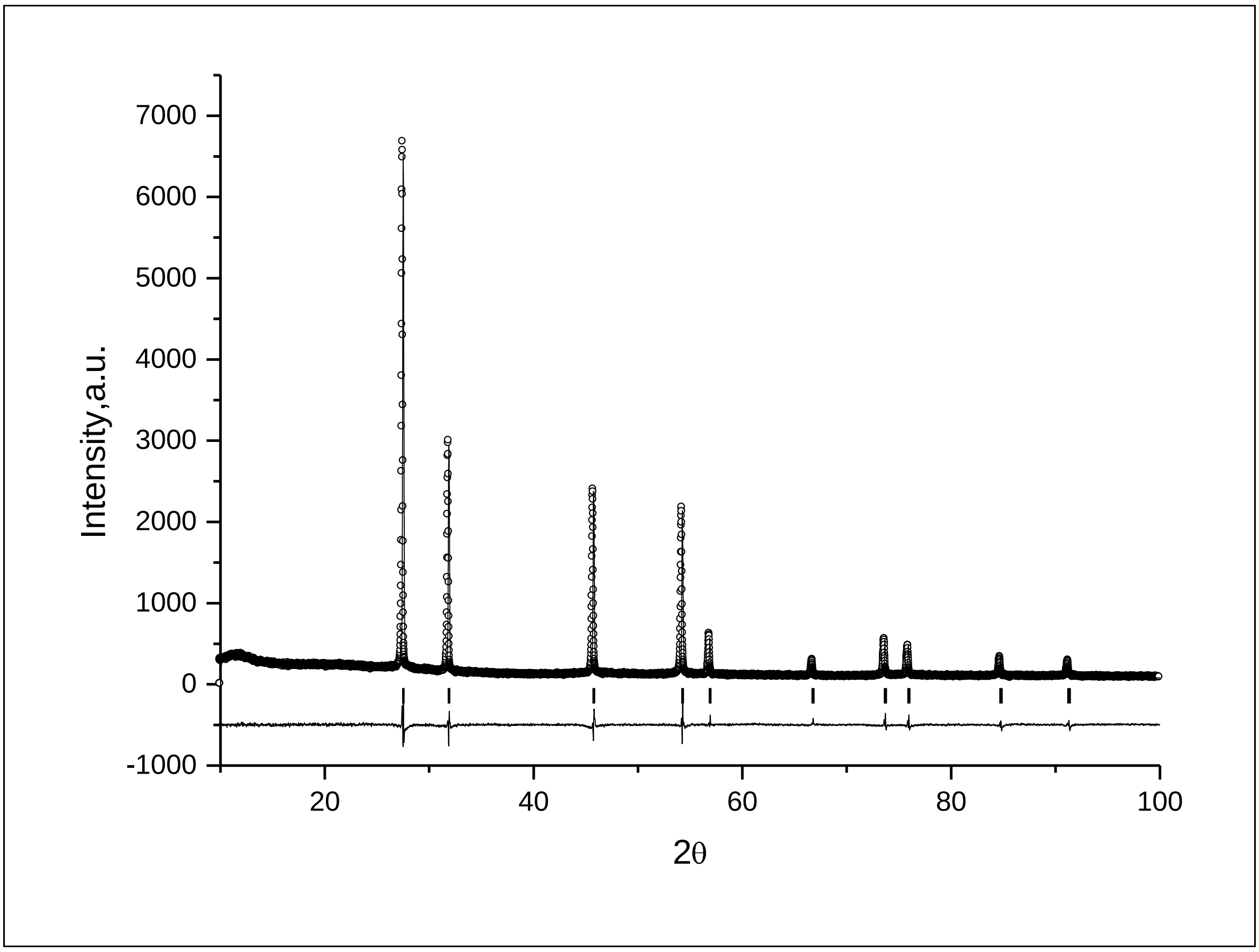

Figure 1. Diffraction pattern of $Bi_{12.5}Sm_{1.5}ReO_{24.5}$

Table 1. Thermochemical data on dissolution enthalpies and formation enthalpies of binary oxides

| Phases | $Bi_2O_3$ | $Sm_2O_3$ | $Re_2O_7$ | Ref. |
|---|---|---|---|---|
| $\Delta_{sol}H°$, kJ/mol* | -114.4 ± 1.1 | -394.0 ± 3.6 | -27.2 ± 0.1 | This work |
| $\Delta_f H°$, kJ/mol | -578.0±0.2 | -1827.1±0.3 | -1271.9 ± 8.4 | [12] |

* All experimental values were calculated as the average value from six parallel experiments. The errors were calculated for 95% confidential interval using standard procedure of treatment of experimental results [14].